# Metallic states in $Pb_{10}(PO_4)_6O$ induced by the Cu/O-insertions and carrier doping

Liang Liu[*], Xue Ren and Jifan Hu[*]

School of Physics, State Key Laboratory for Crystal Materials, Shandong University, Jinan 250100, China

[*]corresponding author: liangliu@mail.sdu.edu.cn and hujf@sdu.edu.cn

# Abstract

The reports of room-temperature superconductivity in Cu-doped Pb-apatite draw intense interests and debates. Herein, based on the density functional theory, we show that the Cu-insertion is thermodynamically stable and further carrier doping can convert the system $CuPb_{10}(PO_4)_6O$ into metal. The metal conduction is mainly along the one-dimensional (1D) Cu-O chains in the c-axis (out-of-plane). The calculated conductance along the c-axis is larger than the in-plane ones by 2 magnitude orders, indicating the 1D conduction behavior. Moreover, the 1D Cu-O chain is an anti-ferromagnetic Mott insulator at zero-doping point due to the super-exchange. Further electron/hole-doping will erase the anti-ferromagnetism. Therefore, the Cu-inserted system $CuPb_{10}(PO_4)_6O$ show transport and magnetic features similar to the cuprate superconductors. On the other hand, the O-insertion can also induce the metallic states, in which the conductance along the out-of-plane direction is higher than the in-plane direction by 6-folds. Our results display the metallization of $Pb_{10}(PO_4)_6O$ via Cu/O-insertions, and suggesting the conductions along the c-axis might dominate the transport behaviors.

# INTRODUCTION

The room temperature superconducting has long been pursuing since the first discovery of superconductive mercury in 1911. The BCS theory indicate that the simple pure metal cannot host superconductivity beyond ~40 K due to the limited strength of electron-phonon couplings.[1] In the mid-1980s, the superconductivity is found in some carrier-doped cuprate beyond 40 K,[2] which is named as high-$T_C$ superconductivity, in which the strong interactions might play a role. Later on, similar high-$T_C$ superconductivity is also found in other transition metal (TM) based compounds,[3] but the highest recorded superconducting temperature is below 140 K. Recently, some hydrogen-based compounds show superconductivity beyond 200 K, which is thought to be the results of significantly strong electron-phonon couplings via hydrogen phonon modes.[4-7] Particularly, the lutetium super-hydride is reported to host superconductivity under 300 K and ~1 GPa,[8] but it was challenged by later study.[9]

Very recently, the Cu-doped lead-apatite $Pb_{10-x}Cu_x(PO_4)_6O$ (0.9<x<1.1) the LK-99 is reported to display superconductivity at room temperature and air condition,[10-11] which draws worldwide interests. However, several experimental and theoretical studies show controversial results. Qiang Hou *et. al.* reported the transport measurements of LK-99.[12] The metallic behavior was observed between 110 K~300 K and the resistance reduced to ~$10^{-5}$ Ω which approaches the accuracy limit of the equipment. Some groups failed to measure the resistance because their LK-99 samples exhibited highly resistive nature.[13-14] Other studies successfully did the transporting measurements but the sample shows semiconductor behaviors below 300 K.[14-16] Particularly, Shilin Zhu *et. al.* synthesized mixture of 30% LK-99 and 70% $Cu_2S$ and the composite displayed reduction in resistivity by 4 magnitude orders at 380 K which is caused by the structural phase transition of $Cu_2S$. It seems to explain the superconducting-like transition at 380 K in the original work of LK-99.[10-11] However, the sample of Shilin Zhu et. al presents the semiconductor behavior contrasting the original reports[10-11] and later reports[12], and the $Cu_2S$ content of 70% is also significantly distinct with the other reports.[14] So, $Cu_2S$ cannot be the full story. Nonetheless, the impurity should play essential roles in the transport property of LK-99. Since the metallicity is the prerequisite for superconductor, how to understand the impurity-induced metallicity is the central issue.

The Meissner effect is the other key feature of superconductor, which prevents the variations of magnetic flux cross the sample. So, the superconductor can move around a permanent magnet in stable orbitals, which is insanely different with the diamagnet that will simply slip away from the top of single permanent magnet. In the original reports of LK-99 and later study,[10-11, 15] the stable but half levitation of sample on one single permanent magnet was observed. In the original works, these phenomena were explained by the Meissner effect of some superconducting parts in LK-99.[10-11] Since the levitations and movements are not as stable as conventional superconductor, the content of superconductor in the LK-99 sample should be very few. Hence, the dominant phase in LK-99 the $Pb_9Cu(PO_4)_6O$ cannot be the source for superconducting. Which impurity phase or impurity-induced phase contribute the Meissner effects? Besides, if the dominant phase is to repel the permanent magnet, there must be other content to be attracted by the permanent magnet to stabilize the observed half-levitation, such as the ferromagnetic or paramagnetic contents. In another study,[15] the researcher indeed detected some signals of ferromagnetism in their LK-99 in the background of diamagnetism. The question is, what is the source for the sufficiently strong ferromagnetism or paramagnetism in LK-99 under room temperature?

Several theoretical investigations were devoted to understand the transport and magnetic behaviors of LK-99. The standard density functional theory (DFT) indicate that the pure Pb-apatite is an insulator with large energy gap ~5 eV, while the Cu-doping can lead to non-dispersive impurity bands in the energy gap.[13, 17-26] By now, all the calculations treat the impurity Cu as substitutions on Pb site as suggested by the original works.[10-11] Since there are two symmetrically inequivalent 4f-Pb and 6h-Pb in $Pb_{10}(PO_4)_6O$, at least two inequivalent $Pb_9Cu(PO_4)_6O$ models are possible *i.e.*, 4f-$CuPb_9(PO_4)_6O$ and 6h-$CuPb_9(PO_4)_6O$. The former case is assumed in the original works,[10-11] which presents zero band gaps on GGA level.[14] However, this metallicity is unstable and the band gap will be open by further interactions such as the strong interactions between d-electrons,[17, 21, 25-26] the phonon-electron-coupling induced lattice distortions,[19] the spin-orbital-couplings,[24] *etc*. The 6h-$CuPb_9(PO_4)_6O$ is energetically favored, and it also presents a large band gap ~0.8 eV.[17] Hence, the Cu-substitutions on Pb always lead to the semiconductor phase, in line with most of the experiments that show the semiconducting in the high-quality LK-99 sample, ruling out the metallicity induced by simple Cu-substitutions. On the other hand, the spin-polarized calculations show that the Cu-substations can induce magnetic moment on Cu ions but the exchange coupling are weak (<0.1 meV).[17-20] So, the system cannot host long-range order at room temperature. There must be other ferromagnetic or paramagnetic parts in the sample to explain the observations of half-levitation of sample on single permanent magnet.[10-11, 15] Thus, which impurity can possibly induce the insulator-metal transition and how the impurity causes paramagnetic or ferromagnetic in $Pb_{10}(PO_4)_6O$ are the two vital questions to uncover the mystery of LK-99 and promote the study on superconductivity.

In this work, we investigated the electronic and magnetic structures of Pb-apatite with Cu/O insertions, as inspired by the experimental progresses. The DFT calculations on the Gibbs free energies show that the Cu-inserted system $CuPb_{10}(PO_4)_6O$ is thermodynamically stable. And the $CuPb_{10}(PO_4)_6O$ is a Mott semiconductor with small gaps and anti-ferromagnetic couplings between Cu ions. Further electron/hole-doping can convert the system into metal, and the conducting states are in the 1D Cu-O chain along the c-axis the out-of-plane direction. The dc conductance in out-of-plane direction is larger than the other directions by two magnitude orders, showing perfect 1D conductions. Moreover, the carrier-doping also suppress the anti-ferromagnetism, revealing many similarities with the high-$T_C$ superconductor of cuprate except the lowered transporting dimension. Besides, the O-inserted system can be stabilized by further Cu-substitutions on Pb. The thermodynamically stable $Pb_9Cu(PO_4)_6O_2$ is also metallic and the conductance are nonzero in all direction but the c-axis conductance is larger than the a/b-axis ones by 6-folds. All these results propose the possible mechanism to induce the metal in Pb-apatite with Cu/O insertions. The exotic 1D conductive Cu-O chain with low-dimensional conductance and carrier-density dependent magnetism may shed light to the further study on the superconductivity.

## RESULTS AND DISSCUSIONS

In the conventional structure of $Pb_{10}(PO_4)_6O$, 25% partially occupied oxygen is located at the origin of the cell presented in Fig. 1a. Hence, there are plenty of space for the interstitial impurities. Here we consider four kinds of impurities: 1. the O-insertion in the center to mimic the over oxidations for the sample annealing in air condition (Fig. 1a); 2. the Cu-insertion in the center to mimic the cation interstitial (Fig. 1b); 3. the O-insertion and the Cu-substituion on 6h site (Fig. 1c); 4. the O-insertion and the Cu-substitution on 4f site (Fig. 1d). Since the LK-99 are prepared in high temperature ~900 K,[10-11] the thermodynamic potentials are good to identify the stability of structures with or without impurities. Due to the importance of $Cu_2S$ and oxygen as indicated by several studies,[27-28] the potentials of $Cu_2S$, PbO (red), $O_2$ and $SO_2$ are took for reference. And the potential of $O_2$ is inferred from the experimental formation heat of red PbO and calculated free energy of Pb metal. For the first case of O-insertion, the free energies are compared with $\Delta G = G(Pb_{10}(PO_4)_6O2) - G(Pb_{10}(PO_4)_6O) - G(O_2)/2$, which gives 1.1 eV. So, when the thermodynamics are sufficiently equilibrated, the possibility to find over-oxidized Pb-apatite is very low. To identify the Cu-insertion, the potential is $\Delta G = G(CuPb_{10}(PO_4)_6O) + G(SO_2)/2 - G(Pb_{10}(PO_4)_6O) - G(O_2)/2 - G(Cu_2S)/2$, giving –0.0 eV. Hence, the Cu-inserted Pb-apatite can be produced by providing sufficient oxygen and $Cu_2S$. The equation for the last two cases is the same: $\Delta G = G(CuPb_{10}(PO_4)_6O_2) + G(SO_2)/2 - G(Pb_{10}(PO_4)_6O) - G(Cu_2S)/2 - G(O_2)$. For the 4h and 6f substitutions, $\Delta G$ is − 0.5 eV and − 0.3 eV, indicating that they are thermodynamically possible when oxygen and $Cu_2S$ are sufficient. So, the Cu-doping can stabilize the over-oxidized Pb-apatite.

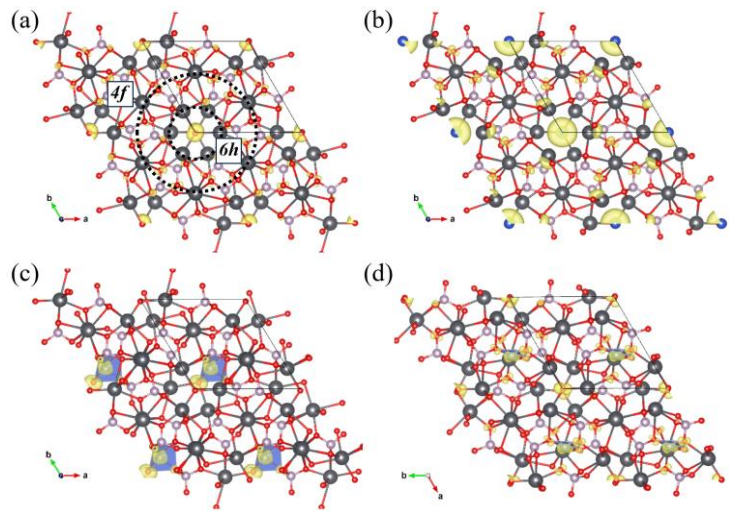

**Figure 1** Geometry structure and spin density. (a) O-inserted apatite $Pb_{10}(PO_4)_6O_2$. (b) Cu-inserted apatite $CuPb_{10}(PO_4)_6O$. (c) O-inserted and Cu-substituted apatite 6h-$CuPb_9(PO_4)_6O_2$. (d) O-inserted and Cu-substituted apatite 4f-$CuPb_9(PO_4)_6O_2$. The grey, blue, red and purple balls denote the Pb, Cu, O and P ions. Blue tetragonal is the Cu-O tetragonal. The small and large dotted circles show the positions of 6h-Pb and 4f-Pb. The yellow surface is the isosurface of spin density with isosurface value 0.01μB/Å$^3$. These are illustrated by VESTA.[29]

All the four models above display the spin-polarization in the ground state. In the O-inserted apatite $Pb_{10}(PO_4)_6O_2$ (Fig. 1a), the two center oxygen ions possess localized magnetic moment of 1 μB, which is the d$^0$ magnetism. For the Cu-inserted case in Fig. 1b, the inserted Cu ion shows localized magnetic moment of 1 μB, indicating that the inserted ion is in Cu$^{2+}$ state with d$^9$ configuration. In the case with O-insertion and 6h-substitution (Fig. 1c), the Cu ion and its nearest center O ions contribute magnetic moments of 1 μB for each. In the last case, the Cu-substitution at 4f position is far from the center O ions (Fig. 1d), and the two center ions and the inserted Cu ion all have 1 μB magnetic moments. The energy advantages due to the spin-polarization for the four cases are 66 meV, 51 meV, 464 meV and 389 meV. These energy advantages due to the spin-polarization are sufficiently higher than the thermal fluctuations of room temperature (26 meV). Therefore, the atomic magnetic moments can be stable at room temperature, acting as the sources for the responses in magnetic fields. The long-range magnetic order will be discussed later.

Fig. 2 shows the band structure and partial density of states (PDOS) of the four models on the GGA+U level. The Fermi level sink into the valence bands in the O-inserted Pb-apatite (Fig. 2a), revealing possible metallicity. The PDOS in Fig. 2e shows that the spin-polarized s-orbitals of Pb and p-orbitals of O dominate the states near Fermi level. Fig. 2b displays the small-gap semiconductor behavior in Cu-inserted Pb-apatite. The two bands near the Fermi level are dispersive which is different from the nondispersive ones in only Cu-substitution case in former study.[17, 21, 25-26] And these two bands are contributed by the d-orbitals of Cu, s-orbitals of Pb, and p-orbitals of O, as indicated in Fig. 2f. For the case of O-insertion and Cu-substitution in 6h site, a metallic band crosses the Fermi level as shown in Fig. 2c, which is dominated by the Pb s-orbitals and O p-orbitals (Fig. 2g). Moreover, the two nondispersive impurity bands dominated by the d-orbitals of Cu are also presented. For the O-insertion but Cu-substitution in 4f site, Fig. 2d and 2h display the metallic bands that are dominated by the Pb s-orbitals and O p-orbitals. All these indicate that the O-insertion can induce the metallicity in Pb-apatite, and the Cu-insertion leads to semiconductor with small bands gaps.

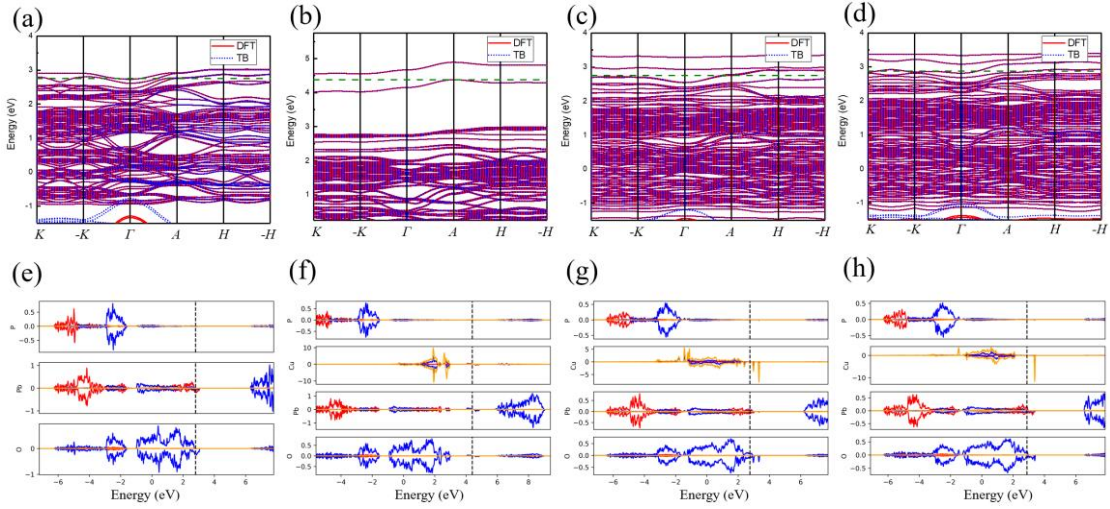

**Figure 2** Band structure and PDOS. (a) Band structure of O-inserted apatite $Pb_{10}(PO_4)_6O_2$. (b) Band structure of Cu-inserted apatite $CuPb_{10}(PO_4)_6O$. (c) Band structure of O-inserted and Cu-substituted apatite $6h$-$CuPb_9(PO_4)_6O_2$. (d) Band structure of O-inserted and Cu-substituted apatite $4f$-$CuPb_9(PO_4)_6O_2$. The red solid lines give the results of DFT, blue dotted lines give the results of Wannier interpolation. The green dashed line denotes the Fermi level. (a) PDOS of O-inserted apatite $Pb_{10}(PO_4)_6O_2$. (b) PDOS of Cu-inserted apatite $CuPb_{10}(PO_4)_6O$. (c) PDOS of O-inserted and Cu-substituted apatite $6h$-$CuPb_9(PO_4)_6O_2$. (d) PDOS of O-inserted and Cu-substituted apatite $4f$-$CuPb_9(PO_4)_6O_2$. Red, Blue and orange lines denote the contributions from s, p and d-orbitals. Positive and negative values denote the contributions of spin-up and spin-down states.

Up to the constant relaxation time approximation, the dc-conductance is approached by the Boltzmann transporting theory. Fig. 3a shows that the O-insertion induced anisotropic conductance and the largest one is along the z-direction with $\sigma^{zz}/\tau \approx 6\times10^{19}$ S/(m·s) at zero carrier-doping. The in-plane dc-conductance are smaller $\sigma^{xx}/\tau \approx \sim 1\times10^{19}$ S/(m·s). Hence, the metallicity induced by O-insertion mostly transport the carriers in z-direction. The effective relaxation time $\tau$ is hard to be determined by first-principle, but sensitive to the quality of the sample. Assuming typical values in metals, says $\tau \approx 10$ fs, we then have $\sigma^{zz} \approx 6\times10^5$ S/m, which is worse than the cooper by two magnitude orders ($\sigma_{Cu} \approx 5.9\times10^7$ S/m) and worse than the Pb metal by one magnitude orders ($\sigma_{Cu} \approx 9.6\times10^6$ S/m).

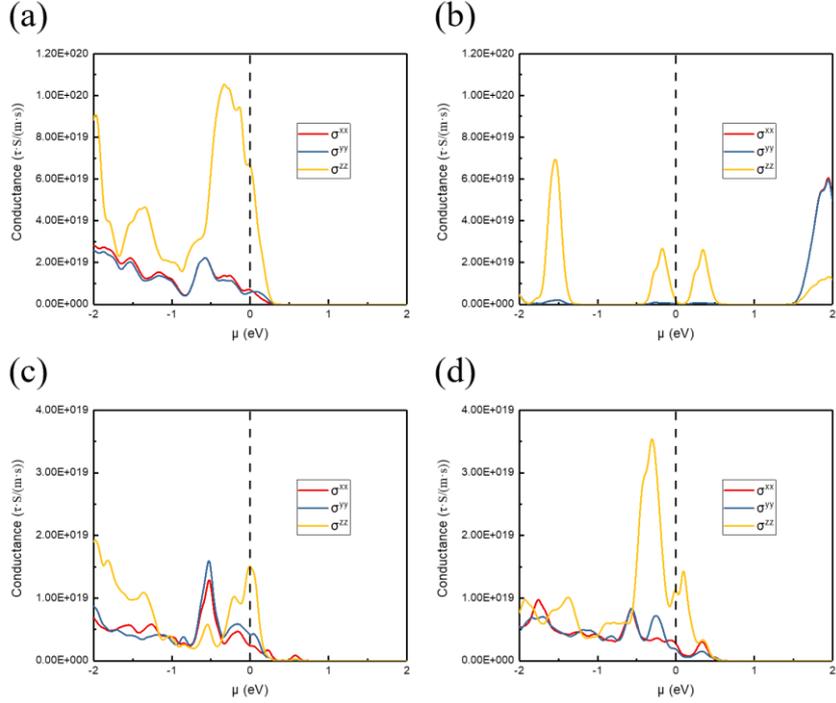

**Figure 3** DC conductance. (a) O-inserted apatite $Pb_{10}(PO_4)_6O_2$. (b) Cu-inserted apatite $CuPb_{10}(PO_4)_6O$. (c) O-inserted and Cu-substituted apatite 6h-$CuPb_9(PO_4)_6O_2$. (d) O-inserted and Cu-substituted apatite 4f-$CuPb_9(PO_4)_6O_2$. Dashed line denotes the neural point without electron or hole doping.

The Cu-inserted Pb-apatite shows negligible dc-conductance at zero carrier-doping (Fig. 3b), in line with the semiconductor nature as shown by the band structures in Fig. 2b. Either electron or hole doping can lead to the nonzero dc-conductance. And the conductions around the neural point are extremely anisotropic. At $\mu = -0.2$ eV, the conductance along $z$-direction is $\sigma^{zz}/\tau \approx 1\times10^{19}$ S/(m·s), larger than the conductance along $x$ and $y$ directions by two magnitude orders. Thus, the conduction route in Cu-inserted system is one-dimensional, implying the role of 1D Cu-O chain. For the system with O-insertion and Cu-substitution, Fig. 3c and 3d show that they also present anisotropic transporting behavior and the easy-conduction routes are along z-direction. All above results indicate that the conduction induced by O and Cu insertions all show better conducting along the out-of-plane direction. Therefore, the transport experiments might be better to measure along the c-axis.

To further study the reason for the significantly anisotropic conducting behavior, the carrier distributions in real-space are investigated. For the O-inserted system, Fig. 4a shows the summed moduli of wavefunctions of all the states near Fermi level by ±0.1 eV, which are the relevant states for transporting. Apparently, the carrier orbitals are dominated by the orbitals of center oxygens and the 6h-Pb, which mainly spreads along the vertical direction, showing alignments with the conductance advantages in out-of-plane direction.

Fig. 4b displays the carrier orbitals of Cu-inserted system, they are the moduli of wavefunctions of states near the electron doping level $\mu = -0.2$ eV. Clearly, the carrier orbitals consist of the orbitals of 6h-Pb, the center oxygen and the inserted Cu. They form a nearly perfect 1D conduction route along the vertical direction, in line with the 1D conductance in Fig. 3b.

Fig. 4c and 4d display the carrier orbitals of O-inserted and Cu-substituted systems. They are the moduli of wavefunctions of states near the Fermi level, show similarities with the O-inserted system in Fig. 4a. The center oxygen and 6h-Pb contribute the most part of the carrier orbitals, leading to the better conduction along the vertical direction.

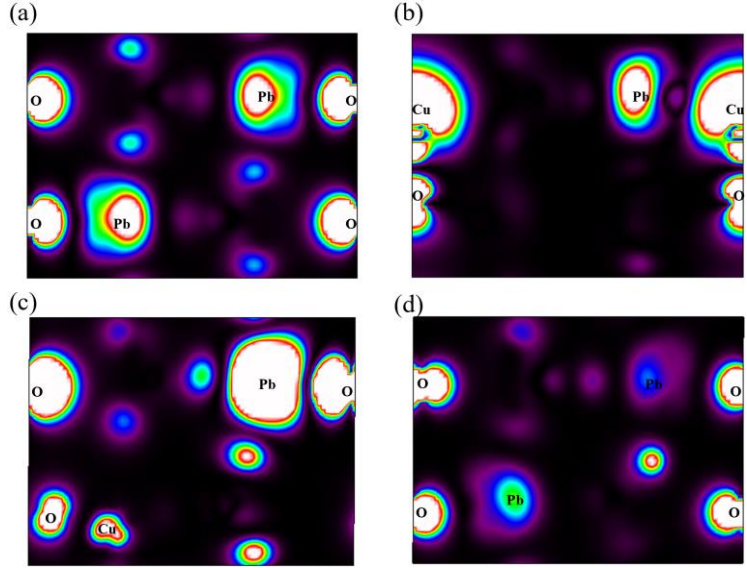

**Figure 4** The moduli of wavefunctions for conducting states. (a) States near Fermi level for O-inserted apatite $Pb_{10}(PO_4)_6O_2$. (b) States below Fermi level by 0.175 eV for Cu-inserted apatite $CuPb_{10}(PO_4)_6O$. (c) States near Fermi level for O-inserted and Cu-substituted apatite 6h-$CuPb_9(PO_4)_6O_2$. (d) States near Fermi level for O-inserted and Cu-substituted apatite 4f-$CuPb_9(PO_4)_6O_2$. These data are in the (1 -1 0) plane cross the cell origin. The black and white regions denote the zero values and moduli greater than 0.002 e/Å$^3$.

The long-range magnetic order depends on the exchange couplings (J) between the magnetic ions. Based on the perturbation theory, the couplings in the four systems are calculated and presented in Fig. 5. For the O-inserted Pb-apatite, the exchange couplings between the localized magnetic moments on center oxygen are mainly ferromagnetic, as shown in Fig. 5a. The dominated exchange couplings are along the in-plane direction with J ≈ −2.6 meV, which is tiny and the ferromagnetic order cannot be established beyond 11 K (see Fig. S3 in supporting information).

For the Cu-inserted system, Fig. 5b shows that the exchange couplings between the Cu ions are dominated by the anti-ferromagnetic interactions. Since the Cu-O-Cu bond angles is 180° along the 1D Cu-O chain, the anti-ferromagnetism is in line with the super-exchange mechanism.[30] Due to the nature of semiconductor in Cu-inserted apatite $CuPb_{10}(PO_4)_6O$, the exchange couplings exponentially die away as the increasing of distance.

Since the metallization of $CuPb_{10}(PO_4)_6O$ needs further carrier doping, it is intriguing to see the dependence of anti-ferromagnetic exchange couplings on the carrier doping. Fig. 6 shows that both the electron and hole doping can significantly alter the exchange couplings between the Cu ions along the Cu-O chain. The negative μ corresponds to the hole-doping, and the anti-ferromagnetic exchange couplings are erased at μ ≈ −0.1 eV and change sign for lower μ. For the positive μ the electron-doping, the anti-ferromagnetic interactions will increase first and then decrease to zero at µ ≈ −0.2 eV. Further electron-doping also leads to the sign change of J.

Fig. 5c shows the exchange couplings in system with O-insertion and Cu-substitution in 6h site. The couplings between the Cu ions and its nearest center O ions are as large as −96 meV, showing the strong tendency that the magnetic moments of Cu and O ion are in parallel. However, the exchange couplings quickly die away at longer distance. So, the long-range order cannot be established in this case but possible paramagnetic may be presented. For the system with O-insertion and Cu-substituion in 4f site, Fig. 5d shows that the exchange couplings between O ions are anti-ferromagnetic, while the ferromagnetic exchange dominate between Cu and O ions. However, the largest coupling energy is as weak as ~1 meV, thus no long-range order can be established.

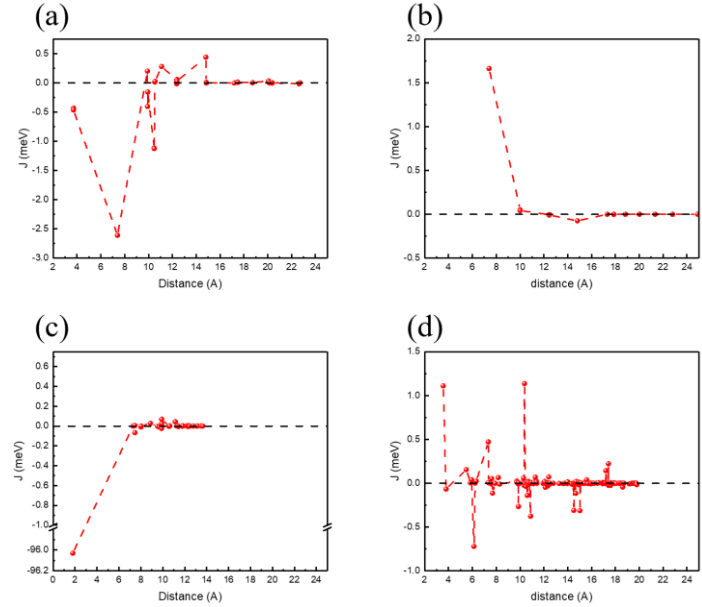

**Figure 5** Exchange coupling strength depends on the distance between magnetic ions. (a) O-inserted apatite $Pb_{10}(PO_4)_6O_2$. (b) Cu-inserted apatite $CuPb_{10}(PO_4)_6O$. (c) O-inserted and Cu-substituted apatite 6h-$CuPb_9(PO_4)_6O_2$. (d) O-inserted and Cu-substituted apatite 4f-$CuPb_9(PO_4)_6O_2$. Positive values correspond to the anti-ferromagnetic type couplings and negative values are ferromagnetic couplings.

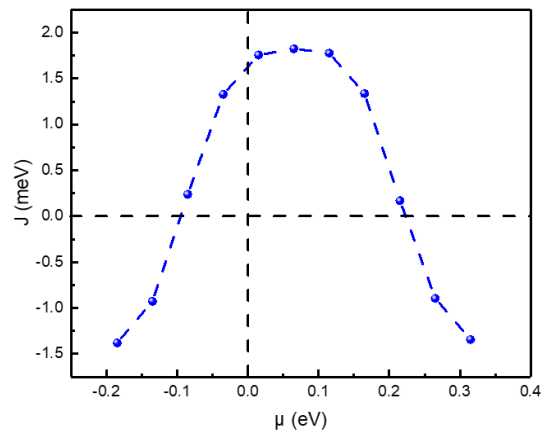

**Figure 6** Exchange couplings between the nearest Cu ions in the Cu-inserted apatite $CuPb_{10}(PO_4)_6O$, which depends on the carrier-doping.

# CONCLUSION

In summary, the thermostability, transport and magnetic properties of the Pb-apatite with four kinds of impurities are investigated. The Cu-insertion is stable and leads to the Mott semiconductor in Pb-apatite and the Cu-O chain is initially antiferromagnetic, in which the Cu ion possesses 1μB magnetic moment. With further carrier-doping, the anti-ferromagnetism is suppressed and Cu-O chain becomes conductive. The conduction shows significantly 1D feature, in which the dc conductance along the Cu-O chain is larger than the orthogonal directions by two magnitude orders. Up to the constant relaxation $\tau \approx 10$ fs, the dc-conductance of Pb-apatite with Cu-O chain can reach $\sim 10^5$ S/m, two magnitude orders smaller than cooper. On the other hand, the system with O-insertions and further Cu-substitutions also show thermal stability and metallicity properties with the highly anisotropic dc conductance. The easy conducting direction is along the out-of-plane $c$-axis and the $\sigma^{zz}$ is larger than the other orthogonal directions by 6 folds. The exchange couplings for the O-insertions and Cu-substitutions are complex and weak, hard to produce long-range magnetic orders. The above results provide the possible explanations for the metallic behavior as observed in some experiments, but also suggesting the Cu-inserted system $CuPb_{10}(PO_4)_6O$ have transport and magnetic properties similar to the cuprate superconductor. Although the results above are far from enough to clarify the superconductor in LK-99, the possible role of impurities effects of Cu/O insertions and especially the 1D Cu-O chain may merit future studies on the system.

# ACKNOWLEDGEMENTS

This work is supported by Natural Science Foundation of Shandong (Grant No. ZR2022QA019), National Natural Science Foundation of China (Grant Nos. 12074221, 52171181, 52002222, 2021-869, 11904204, 51472150), and Natural Science Foundation (No. 2022A1515011667) of Guangdong.

# Notes

The authors have no conflicts to disclose.

# Supporting Information

# One-dimensional conduction routes in $Pb_{10}Cu(PO_4)_6O$ induced by the Cu/O-insertions

Liang Liu[*], Xue Ren and Jifan Hu[*]

School of Physics, State Key Laboratory for Crystal Materials, Shandong University, Jinan 250100, China

*corresponding author: liangliu@mail.sdu.edu.cn and hujf@sdu.edu.cn

# Computation details

The calculations of density functional theory are done with the projector augmented plane-wave basis, which is implemented in Vienna *ab-initio* simulation package.[1-2] And the plane-waves are cut-off at 550 eV. The exchange-correlations of electrons are described by the generalized gradient approximations with the form proposed by Perdew, Burke, and Ernzerhof[3]. To improve the descriptions on the d-electrons, Hubbard $U$ with $U_{eff}$ = 2.0 eV is applied on the d-shell of Cu ions. The energy converge criterion for solving self-consistent Kohn-Sham equations is $10^{-5}$ eV. The Brillouin zone is sampled with resolutions better than 0.02 Å$^{-1}$, using the scheme of Monkhorst-Pack[4]. The structures in this study are fully relaxed until the Hellman-Feynman smaller than 0.01 eV/Å. The real-space Hamiltonian of ground state is interpolated with Wannier functions as implemented in WANNIER90 package[5-6].

The dc conductance are computed via the Boltzmann transporting theorem up to the constant relaxation time approximations:

$$\sigma^{\alpha\beta} = -\tau e^2 \sum_k \int dE\, v_n^\alpha v_n^\beta\, \partial_E f(E_n - \mu) \quad (S1)$$

in which $\tau$ is the relaxation time, $v_n^\alpha$ is the diagonal element of velocity operator and $f(E_n - \mu)$ corresponding to the Fermi-Dirac distributions. The electron temperature $T$ is set to 298.15 K and the BZ-integration for eq. S1 are on the resolution of 0.01 Å$^{-1}$.

The magnetic coupling energy between ions α and β are computed via the magnetic force theorem:[7]

$$J_{\alpha\beta} = \frac{1}{4\pi}\, Im \int_{-\infty}^{E_f} dE\, Tr^L\{\Delta_\alpha G_{\alpha\beta}^\uparrow \Delta_\beta G_{\beta\alpha}^\downarrow\} \quad (S2)$$

in which $\Delta$ is the exchange splitting and $G^{\uparrow(\downarrow)}(E) = (E - H^{\uparrow(\downarrow)} + i\eta)^{-1}$ is the Green's function of spin-up (spin-down) electrons. $H^{\uparrow(\downarrow)}$ is the real-space Hamiltonian of spin-up (spin-down) electrons, and $\eta$ is set to 0.0001 eV.

The temperature dependent evolutions of magnetizations and the magnetic order temperature are obtained by the Hatree-Fock spin-wave theory[8] and the Monte Carlo simulations on Heisenberg spin model as implemented in MCSOLVER.[9-10] 4×10$^6$ Metropolis sampling sweeps were used for thermal equilibrations and subsequent 8×10$^6$ sweeps were conducted for the statistics on thermal quantities.

# Structural, electronic and magnetic properties of Pb$_9$Cu(PO$_4$)$_6$O

Fig. S1 shows the geometry of Cu-substitutions on 4f-Pb and 6h-Pb. Note that the choosing of particular occupation of oxygen in the center can lower the symmetry. There are actually 6 inequivalent Pb position per one cell. The structures shown in Fig. S1 correspond to the energetically favored ones for the two kinds of substitutions. Further, the ΔG for the 6h-subsitution is lower than the 4f case by 0.6 eV, indicating the advantages of 6h-substitutions, in line with former study. [11]

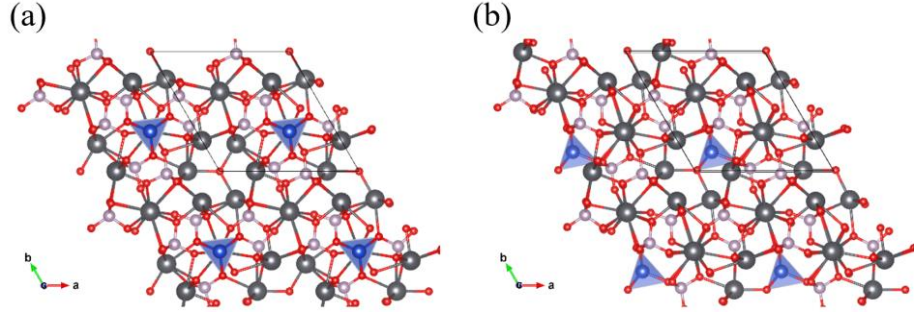

**Fig. S1** The atomic structure of (a) 4f-$CuPb_9(PO_4)_6O$ and (b) 6h-$CuPb_9(PO_4)_6O$.

Fig. S2 shows the electronic structures for the $CuPb_9(PO_4)_6O$. The band structure of 4f-$CuPb_9(PO_4)_6O$ in Fig. S2a presents two nondispersive bands, corresponding to the impurity levels of Cu-d states. These flat bands cannot contribute to the conduction due to the decoupling to other environment states as shown in the PDOS. Fig. S2c displays the band structure of 6h-$CuPb_9(PO_4)_6O$, revealing a semiconductor with band gap ~0.7 eV (on the level of GGA+$U$, and $U_{eff,Cu:d}$=2.0).

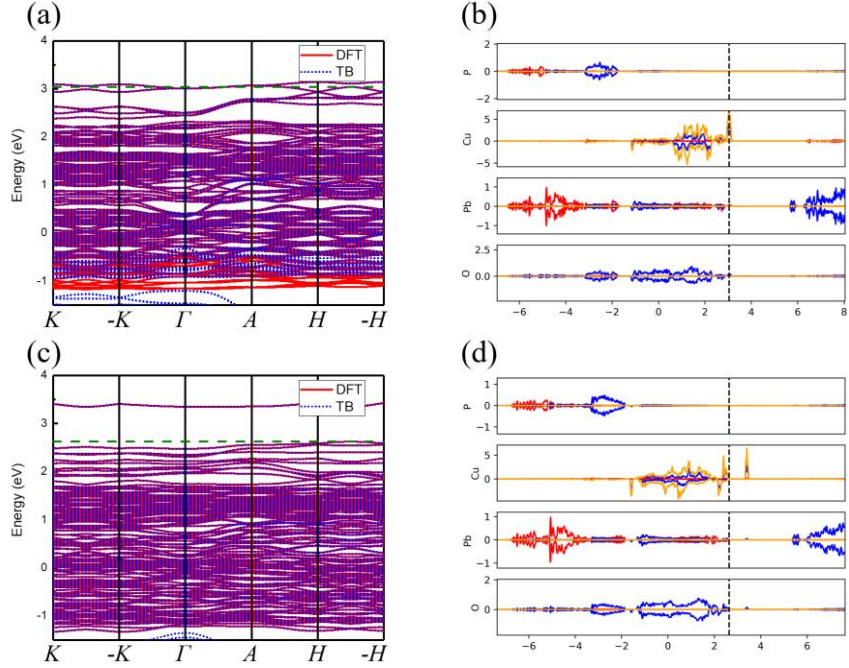

**Fig. S2** Band structures and partial density of states. (a) Band structure and (b) PDOS of of 4f-$CuPb_9(PO_4)_6O$. (c) Band structure and (d) PDOS of 6h-$CuPb_9(PO_4)_6O$.

The exchange couplings for $CuPb_9(PO_4)_6O$ are shown in Table S1. The 4f-substitution can lead to the ferromagnetic couplings but the sub-meV strength cannot support ferromagnetism beyond 3 K, as indicated by the Monte Carlo simulations displayed in Fig. S3. For the 6h-substitution, the couplings are weak, too. No long-range magnetic order can be expected.

**Table S1** Exchange coupling energies between Cu ions in $CuPb_9(PO_4)_6O$ in unit meV. And the transition temperature for long-range magnetic order.

|  | J (0 0 1) | J (1 1 0) | J (1 0 0) | J (0 1 0) | Tc |
|---|---|---|---|---|---|
| 4f-$CuPb_9(PO_4)_6O$ | -0.01 | -0.15 | -0.15 | -0.15 | <3 K |
| 6h-CuPb$_9$(PO$_4$)$_6$O | 0.00 | 0.00 | 0.019 | 0.00 | 0 K |

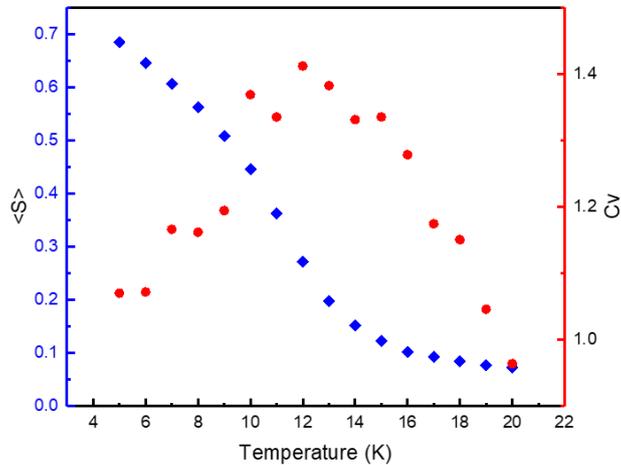

**Fig. S3** The thermally averaged spin and specific heat per cell for the O-inserted Pb$_{10}$(PO$_4$)$_6$O$_2$.

Structural, Electronic, Magnetic Properties of Cu-Doped Lead-Apatite. *ArXiv* 2308.04344 **2023**.